\documentclass[reprint,superscriptaddress,  amsmath,amssymb, aps,floatfix]{revtex4-1}%
\usepackage{graphicx}
\usepackage{dcolumn}
\usepackage{bm}
\usepackage{amsmath}
\usepackage{amsfonts}
\usepackage{amssymb}%
\usepackage{xcolor}
\usepackage{appendix}

\usepackage{MnSymbol}
\usepackage{natbib}

\usepackage[author=Guillermo,color=yellow]{pdfcomment}

\newcommand{\affvqcc}{Vigo Quantum Communication Center, University of Vigo, Vigo E-{36310}, Spain}

\newcommand{\affuvigo}{Escuela de Ingeniería de Telecomunicación, Department of Signal Theory and Communications, University of Vigo, Vigo E-36310, Spain}

\newcommand{\affatlantic}{atlanTTic Research Center, University of Vigo, Vigo E-36310, Spain}

\newcommand{\afftoyama}{Faculty of Engineering, University of Toyama, Gofuku 3190, Toyama 930-8555, Japan}

\begin{document}

\title{Deterministic QKD source robust against side-channel attacks}

\author{Kiyoshi Tamaki}
	\affiliation{\afftoyama}
    \author{Marcos Curty}
	\affiliation{\affvqcc} \affiliation{\affuvigo} \affiliation{\affatlantic}  \author{Akihiro Mizutani}
	\affiliation{\afftoyama}
    \author{\'{A}lvaro Navarrete}
	\affiliation{\afftoyama}\affiliation{\affvqcc} \affiliation{\affuvigo} \affiliation{\affatlantic}

\begin{abstract}
Quantum key distribution (QKD) is secure in principle, but practical security can be undermined by discrepancies between real devices and the idealized models assumed in security proofs. Source side channels, including those exploited by Trojan-horse attacks, are particularly detrimental: neglecting them compromises implementation security, whereas accounting for them reduces performance. Here we propose a QKD source that is intrinsically robust against side-channel attacks. Unlike existing passive and modulator-free schemes, it requires neither post-selection of the emitted pulses nor devices with a perfect extinction ratio to suppress side channels, and it does not introduce correlations between the intensity and the encoded bit or basis. Consequently, under idealized source assumptions commonly adopted in proposals for existing schemes, conventional decoy-state security analyses apply directly, yielding substantially higher key rates. Our proposal appears to be within reach of current technology and therefore provides a clear and practical path toward implementation-secure QKD.
\end{abstract}

\maketitle

\section{Introduction}

Quantum key distribution (QKD) enables two distant users, Alice and Bob, to establish a common secret key with information-theoretic security \cite{BennettBrassard1984,LoCurtyTamaki2014}. Security proofs, however, rest on idealized device models, while real-world transmitters and receivers exhibit calibration errors, finite isolation and the encoding of information in uncontrolled modes, inter alia. If not properly accounted for, these imperfections may open side channels through which an adversary (Eve) could acquire information unavailable in the idealized protocol description \cite{LoCurtyTamaki2014}.

Measurement-device-independent (MDI) QKD and related architectures with an untrusted measurement node eliminate all detector side channels \cite{LoCurtyQi2012,RubenokEtAl2013MDI, Lucamarini2018}, thereby bringing transmitter security to the forefront. Transmitter side channels may arise passively, when the emitted light carries setting-dependent information in uncontrolled degrees of freedom, or actively, through Trojan-horse attacks (THAs), in which Eve injects light into Alice's transmitter and analyzes the reflected field \cite{Vakhitov2001,Gisin2006,Jain2014,Sajeed2017,Lucamarini2015,TamakiCurtyLucamarini2016LeakySource}. Security frameworks incorporating such imperfections exist \cite{GLLP2004,TamakiCurtyLucamarini2016LeakySource,PereiraCurtyTamaki2019FlawedLeaky,PereiraEtal2020correlation,Curras-Lorenzo25Optica, CurrasLorenzoEtAl2026CorrelatedEncoders,CurrasLorenzoEtAl2026PhaseErrorCorrelatedSources, navarrete2026numerical,sixto2025quantum}, but they require partial experimental characterization of the leakage, and they generally reduce the key rate. Experimental countermeasures such as filters, isolators, attenuators, and monitoring detectors provide only finite, implementation-dependent protection \cite{Gisin2006,Jain2015Risk,Dixon2017Countermeasures,watchdog-attack2015}.

These concerns have motivated the development of passive sources and modulator-free transmitters, which avoid conventional external optical modulators by assigning the settings via post-selection based on local measurement outcomes \cite{CurtyMoroderMaLutkenhaus2009,CurtyMaQiMoroder2010,CurtyMaLoLutkenhaus2010,ZapateroWangCurty2023Passive,WangEtAl2023FullyPassive,ZapateroCurty2024FiniteKey,LuEtAl2023Passive,HuEtAl2023Passive}, or via electrical laser driving, optical injection locking or interference, or path selection \cite{LoEtAl2023ModulatorFree,Gnanapandithan2026Mitigating}. These approaches, however, have important drawbacks: post-selection reduces the usable signal fraction and secret-key rate \cite{WangEtAl2023FullyPassive,ZapateroWangCurty2023Passive,LuEtAl2023Passive,HuEtAl2023Passive,ZapateroCurty2024FiniteKey}; the intensity modulator often used to discard undesired pulses has a finite extinction ratio, so the generated pulses contain residual side channels \cite{LuEtAl2023Passive,HuEtAl2023Passive,LoEtAl2023ModulatorFree,Gnanapandithan2026Mitigating,NavarreteZapateroCurty2025ModulatorFree}; and the emitted intensity may be correlated with the encoded bit or basis, requiring dedicated security proofs \cite{WangEtAl2023FullyPassive,ZapateroWangCurty2023Passive,ZapateroCurty2024FiniteKey}.

In this work, we propose a QKD source that is robust against side-channel attacks, including THAs. It requires neither post-selection of the emitted pulses nor devices with a perfect extinction ratio to eliminate side channels. Also, it does not introduce correlations between the intensity and the encoded bit or basis. Although the source employs active phase modulators, the applied phase shifts are masked by random phases unknown to Eve, and therefore even perfect knowledge of these phase shifts does not reveal any setting information to her. This enables the use of simpler security proofs \cite{TamakiEtAl2014LossTolerant,PhysRevA.102.062607,WangEtAl2009SourceErrorsStatisticalFluctuations} that do not explicitly account for side channels or intensity–basis-bit correlations, thereby significantly improving performance. Moreover, the source is compatible with a broad range of decoy-state BB84-type protocols in both prepare-and-measure and measurement-device-independent (MDI) configurations.

The main experimental challenges are interfering pulses from two independent gain-switched lasers, precisely measuring their relative phase, and applying accurate feed-forward phase shifts. Importantly, inaccuracies in these processes do not create side channels; they merely provoke state preparation flaws and intensity fluctuations, which increase the quantum bit error rate in the qubit space, but whose impact on performance is significantly less than that of side channels \cite{TamakiEtAl2014LossTolerant}. Moreover, the required interference and high-precision (milliradian-level) phase-control technologies have already been demonstrated \cite{Yuan2014interference, Comandar2016,KeysightM8194A,Drayss2025OAWG,Sinclair2018OpticalOscillators}, making our proposal a promising short- to medium-term route toward implementation-secure communication in future networks.

\section{Results}

\subsection{Assumptions and Source architecture}
We begin by introducing the key components of our source and the assumptions concerning them.

\begin{enumerate}

\item[(A1)] The optical modes emitted by multiple gain-switched laser sources (GSLs) overlap perfectly, and the intensities of the output pulses are stable.

\item[(A2)] The phase of each emitted pulse from the GSLs is randomized independently of the phases of all other pulses, and the phase information is inaccessible to Eve.

\item[(A3)]
A relative-phase measurement (RPM) can be performed on two bright laser pulses. The measurement outcome is assumed to remain unknown to Eve.

\item[(A4)]
A phase modulator (PM) applies a phase shift determined by the RPM outcome in (A3) and Alice's setting choice. Eve is conservatively assumed to have complete access to the applied shift value.

\item[(A5)]
The electrical signals driving the above devices do not leak information about Alice's setting choices.
\end{enumerate}

Assumptions (A1) and (A2) are commonly adopted in proposals for fully passive transmitters and modulator-free sources and are therefore not specific to our scheme \cite{WangEtAl2023FullyPassive,ZapateroWangCurty2023Passive,ZapateroCurty2024FiniteKey}. To mitigate potential correlations between the global phases of successive pulses, one could use multiple independent GSLs and temporally interleave the resulting pulse trains. For instance, these trains can be combined into a single fiber by optical time-division multiplexing, using calibrated optical delay lines followed by a passive optical coupler~\cite{Liu2024TDMCovert,Aboketaf2010OTDM}. Because information is not encoded in the choice of laser, even if Eve could distinguish which laser source is used each time, this would not pose a security issue. Assumption (A4), on the other hand, represents a worst-case scenario in which Eve can read out the applied phase shift perfectly through a THA. Assumption (A3), together with Assumptions (A2) and (A4), enables the encoding of the setting information while ensuring that the encoded information remains secret from Eve. Finally, Assumption (A5) concerns the security of the classical electronic hardware and is routinely made in essentially all QKD implementations, including device-independent QKD \cite{Curty2019Foiling}. From a technological perspective, Assumptions (A1)--(A3) may be challenging to satisfy, particularly at repetition rates in the GHz regime or higher. Nevertheless, these challenges are almost entirely classical in nature.

\begin{figure}
\includegraphics[scale=0.3]{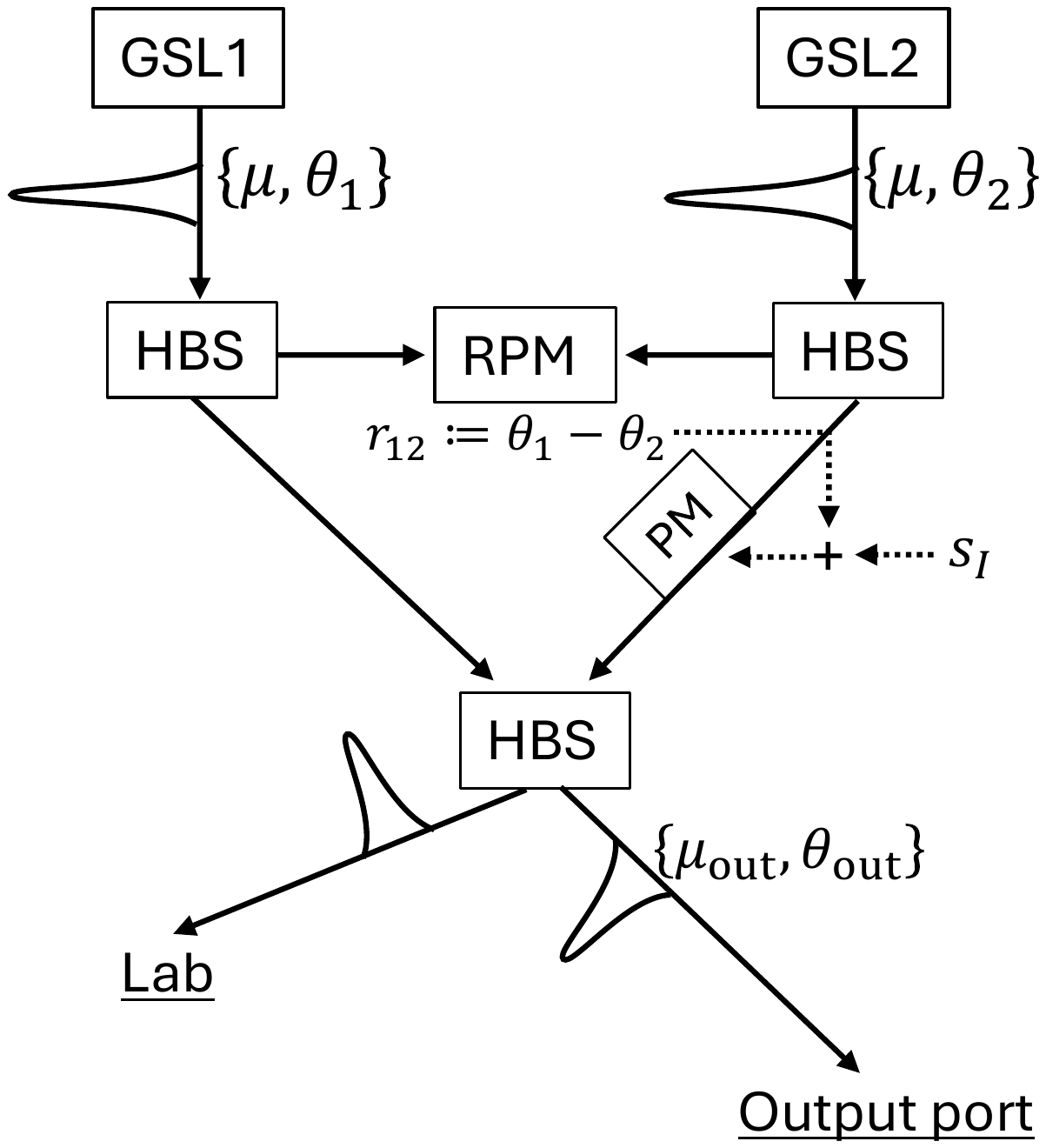}
\caption{Schematic of the first stage, the Intensity Package (IntP). The quantities $\mu$ and $\theta_1$ ($\theta_2$) denote the mean photon number and global phase, respectively, of the pulse generated by GSL1 (GSL2). HBS, RPM, PM, and $s_I$ denote a half-beam splitter, a relative-phase measurement unit, a phase modulator, and the intensity setting selected by Alice, respectively. Specifically, the PM applies the sum of the relative phase $r_{12}$ between the two GSLs, as measured by the RPM, and an additional phase shift $s_I$, chosen according to the desired output intensity, thereby generating an output pulse with the target mean photon number. Importantly, from Eve's perspective, the applied phase $r_{12}+s_I$ is independent of $s_I$, provided that the uniformly random value $r_{12}$ remains secret from her, thereby defeating THAs targeting the PM.}
\label{Fig1}
\end{figure}

Our source consists of two stages. The first stage, shown in Fig.~1 and referred to as the Intensity Package (IntP), generates a phase-randomized coherent pulse with a desired mean photon number. Subsequently, two pulses generated by a single IntP in different time slots are fed into the second stage shown in Fig.~2. This second stage sets the relative phase between the two pulses while preserving global-phase randomization, thereby generating the double-pulse state required for the decoy-state BB84 protocol with phase or time-bin encoding, which can also be converted to polarization encoding \cite{phaseMDI}. We note that the input and output pulses of the IntP, as well as the input pulses to the second stage, are bright, which facilitates high-precision relative-phase measurements. The output pulse pair generated immediately before transmission, however, is attenuated to the single-photon level using an attenuator (Att) and is therefore suitable for QKD. We note that the Att is not intended to eliminate side channels, but rather just to reduce the intensity of the signal pulses. Therefore, it does not introduce security issues given that it applies the attenuation prescribed. To protect the Att against laser-damage attacks, protective components—such as filters, isolators, circulators, or an optical fuse—may be placed downstream of it \cite{Ponosova2022Protecting}. Alternatively, the Att can be omitted by replacing the half-beam splitter (HBS) preceding the PM with an asymmetric beam splitter (see Fig.~2). Note that, although the schemes are depicted using HBSs for simplicity, some of these HBSs may instead be asymmetric.

\subsection{Deterministic intensity preparation and state encoding}
In the IntP (Fig.~1), one pulse from each GSL is split by a half-beam splitter (HBS), and two of the resulting pulses are directed to the RPM to measure the relative phase $r_{12} := \theta_1-\theta_2$, where $\theta_1$ and $\theta_2$ denote the phases of two pulses from GSL1 and GSL2, respectively. For a desired output intensity $\mu_{\rm out}$, Alice selects an additional phase shift---see Eq.~(\ref{outputint}) for the explicit value---and applies the phase shift $r_{12}+s_I$ to one of branches. The phase-modulated pulse and the other pulse then interfere at a final HBS. One output port provides the IntP output, while the other remains inside Alice's laboratory, which could be used to check the quality of the resulting states, or be simply discarded. The mean photon number of the output pulse is given by

\begin{eqnarray}
\mu_{\rm out}
=
\eta_{\rm path}\frac{\mu}{4}
\left|
1
+
e^{is_I}
\right|^2\,,
\label{outputint}
\end{eqnarray}
where $\mu$ denotes the mean photon number of each input pulse and $\eta_{\rm path}$ is the transmittance of the optical-path. Importantly, $\mu_{\rm out}$ is determined by the parameters $s_{I}$, $\mu$, and $\eta_{\rm path}$, which are under Alice's control. The corresponding random phase is given by
\begin{eqnarray}
\theta_{\rm out}={\rm Arg}
\left[
e^{i(\theta_1+\tau)}
+
e^{i(\theta_1+\tau+s_I)}
\right]\,,
\end{eqnarray}
where $\tau$ is the phase accumulated during propagation from the output port of the GSLs to just after the final HBS. Importantly, the phase $\theta_{\rm out}$ depends on $\theta_1$, which is random and unknown to Eve due to Assumptions (A1) and (A2).

\begin{figure}
\includegraphics[scale=0.6]{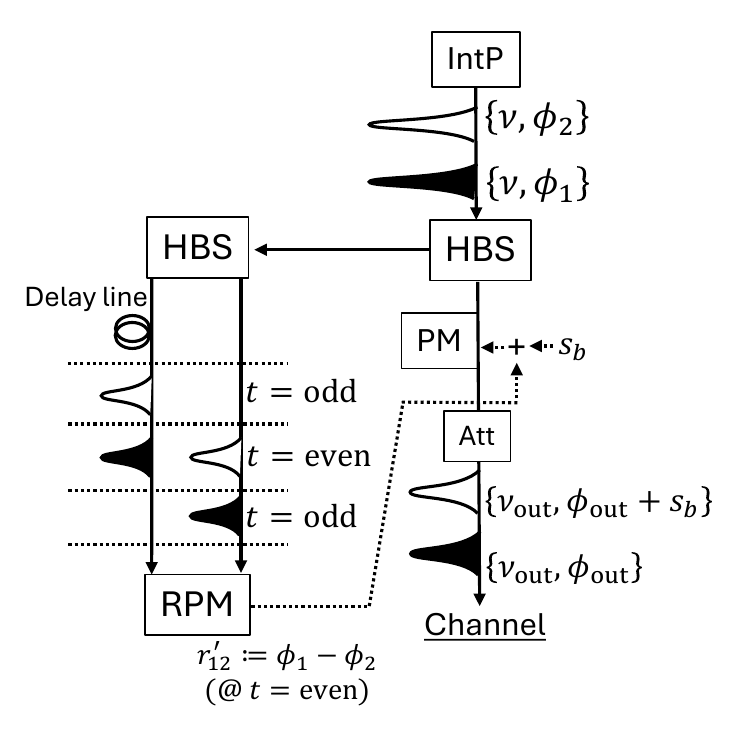}
\caption{Schematic of the second stage. The inputs to this setup are two pulses generated by the IntP and separated in time. The early and late pulses, marked in black and white, respectively, each have mean photon number $\nu$, and global phase $\phi_1$ and $\phi_2$, respectively. These pulses are split by a HBS. One of the resulting pulse pairs is sent to another HBS equipped with a delay line at one of its output ports. This delay line is adjusted so that the pulse originating from the early input pulse overlaps well in time with the late pulse in a time slot referred to as ``even''. The relative-phase measurement (RPM) is then performed only in the ``even'' time slot to measure the relative phase $r_{12}':=\phi_1-\phi_2$. Next, a phase shift of $r_{12}'+s_b$, where $s_{b}$ is selected according to the chosen bit and basis values, is applied only to the late pulse in the other pair of split pulses. After attenuation by the attenuator (Att), the resulting state, which corresponds exactly to a decoy-state BB84 signal state, is sent through the quantum channel. The attenuator can be omitted if the first HBS is replaced with an asymmetric beam splitter. Importantly, from Eve's perspective, the applied phase $r_{12}'+s_b$ is independent of $s_b$, provided that the uniformly random value $r_{12}'$ remains secret from her, thereby defeating THAs targeting the PM.}
\label{Fig2}
\end{figure}

Next, consider two consecutive pulses generated by the IntP. These pulses ideally have identical mean photon numbers, are separated in time, and possess a random relative phase $r_{12}'$. They serve as inputs to the second stage shown in Fig.~2. This stage is conceptually similar to the IntP, except that the phase shift $s_I$ in Fig.~1 is now replaced with $s_b$, which depends on Alice's basis and bit choices rather than on the desired mean photon number. Consequently, applying a phase shift of $r_{12}' + s_b$ to only one of the two pulses sets the relative phase between them to $s_b$. Finally, an Att reduces the intensity of the pulse pair to the single-photon level before it is transmitted through the quantum channel. Assuming a mean photon number $\nu$ for each input pulse and an attenuator transmittance $\eta$, the output mean photon number is given by
\begin{eqnarray}
\nu_{\rm out}
=
\frac{\eta \nu}{2}\,,
\end{eqnarray}
where optical-path losses have been incorporated into $\eta$.
Importantly, the global phase $\phi_{\rm out}=\phi_{1}+\tau'$ of the pulse pair, where $\tau'$ is the phase accumulated during propagation from the input port to the output port of the second stage, is random and unknown to Eve due to Assumption (A2). Therefore, the output state corresponds exactly to the state required for decoy-state BB84.

\subsection{Optical privacy}

The optical privacy of our scheme follows directly from the above construction and the stated assumptions. In both stages, Eve cannot obtain any information about $s_I$ or $s_b$, even if she has complete access to the phase-shift values $r_{12}+s_I$ and $r_{12}'+s_b$. As already mentioned, this is because the random values $r_{12}$ and $r_{12}'$ remain unknown to Eve due to Assumptions (A1)-(A3). Consequently, even if Eve reads out the applied phase shifts successfully through a THA, the obtained information is statistically independent of Alice's encoding settings.
Moreover, our scheme transmits no unused optical signals through the channel, thereby eliminating the possibility that such signals become a side channel. Together with Assumption (A5)---that the electronic driving signals leak no information about Alice's settings---this establishes the optical privacy of the proposed source against side-channel attacks. We emphasize that, thanks to the use of phase shifts, our scheme requires no pulse post-selection and is therefore deterministic. Furthermore, by construction, there are no correlations between the intensity and the encoded bit or basis because these quantities are encoded through independent processes. 

\section{Discussion}

We have proposed a deterministic QKD source for decoy-state protocols---including decoy-state BB84 and MDI-QKD---that effectively defeats Trojan-horse attacks (THAs). The privacy of the source is achieved because the active phase shifts used to generate the states are independent of Alice's setting choices and no unused optical pulses are sent through the channel. 

Provided that the electrical signals driving the components do not leak information about Alice's settings—an assumption common to essentially all QKD implementations—our scheme suppresses optical side channels and renders THAs ineffective. Moreover, the random phase of each gain-switched pulse, together with the random phase applied by phase modulators, ideally remove any correlation between the intensity and the bit-and-basis encoding of the emitted pulses. This means that security proofs that do not explicitly account for side channels or such correlations apply directly, with a substantial gain in performance.

The price to achieve this is experimental, and it is mainly twofold: the pulses from two independent gain-switched lasers must be prepared in perfectly overlapping optical modes, which is experimentally demanding but is becoming feasible with demonstrated technology \cite{Yuan2014interference,Comandar2016}, and the relative phase of each pulse pair must be measured and compensated by a feed-forward phase shift with high precision. Importantly, once Assumption (A1) is satisfied, errors in the relative-phase measurement and feed-forward phase application manifest as state-preparation flaws, such as intensity fluctuations and distorted encoded qubits, rather than as additional optical side channels. This distinction matters because side channels can be substantially more damaging: by enlarging the space spanned by the emitted signals, they may render the four BB84 states linearly independent, exposing the protocol to unambiguous-state-discrimination attacks [47]. Thus, under Assumption (A1), our scheme converts imperfections in interference and phase control into performance penalties rather than additional side-channel leakage. Deviations from perfect mode overlap, on the contrary, would constitute side channels and would have to be treated separately. For implementations in which matching two independent lasers is impractical, we propose in Appendix A a single-laser variant that substantially relaxes this requirement.

Another key assumption of our proposal, which is also widely adopted in most passive and modulator-free source designs, is global-phase randomization of the laser sources. Correlations between global phases become particularly problematic in high-speed systems. To mitigate such correlations, one could combine multiple laser sources using optical time-division multiplexing. This approach could substantially reduce their impact, albeit at the cost of experimental complexity.

\section{Acknowledgments}

K.T. acknowledges support from JSPS KAKENHI Grant Numbers 23K25793 and 23H01096. M.C. acknowledges support from the Galician Regional Government (consolidation of Research Units: AtlantTIC); the Spanish Ministry of Science, Innovation and Universities (MICIU); the Fondo Europeo de Desarrollo Regional (FEDER) through the grant No. PID2024-162270OB-I00; the “Hub Nacional de Excelencia en Comunicaciones Cuánticas” funded by the Ministerio para la Transformación Digital y de la Función Pública and the European Union NextGenerationEU; the European Union’s Horizon Europe Framework Programme under the Marie Sklodowska-Curie Grant No. 101072637 (Project QSI); the project “Quantum Secure Networks Partnership” (QSNP, grant agreement No 101114043); the European Union under the Project IberianQCI (grant 101249593), and the Programa de Cooperación Interreg VI-A España-Portugal (POCTEP) 2021-2027 through the project QUANTUM IBERIA.  A.M. was partially supported by JSPS KAKENHI Grant No. JP24K16977. A.N. acknowledges financial support from the Xunta de Galicia (Conseller{\'i}a de Educac{\'i}on, Ciencia, Universidades e Formac{\'i}on Profesional)
through a Xunta de Galicia Postdoctoral Fellowship (No. ED481B-2025/113).

\appendix

\section{Alternative single-laser implementation}

\begin{figure}
\begin{center}
\includegraphics[scale=0.55]{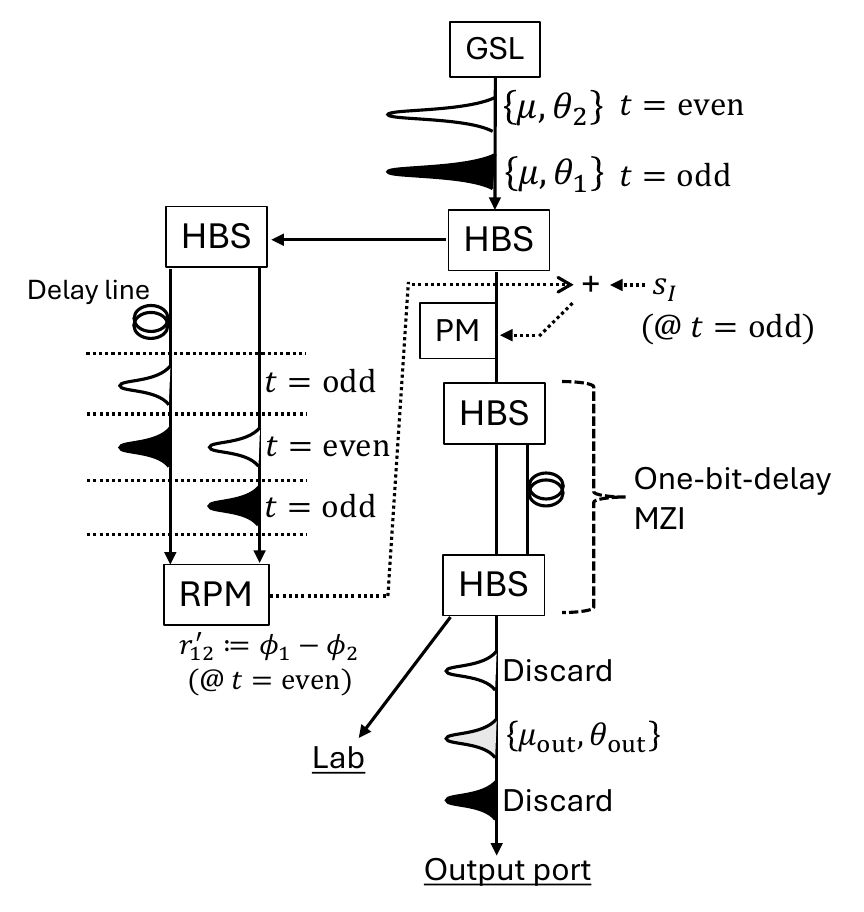}
\end{center}
\caption{Schematic of the alternative setup for the IntP. The middle pulse (marked in gray) at the output port is used as the input to the second stage in Fig.~2. The scheme for relative-phase measurement (RPM) and phase shift is the same as that in Fig.~2, except that the phase-shift value is chosen according to the intended intensity. After applying the phase shift, a one-bit-delay MZI causes the two pulses to interfere, producing three output pulses; the middle pulse is the intended one. The first and last pulses, marked in black and white, respectively, are not used by Alice but carry information about the random phases and must therefore be blocked.}\label{Fig3}
\end{figure}

The scheme introduced in the main text requires two GSLs. Alternatively, one could employ only one laser to increase the interference visibility, but it comes at the cost of introducing side-channel leakage due to the finite extinction ratio of practical devices \cite{NavarreteZapateroCurty2025ModulatorFree}. 

The basic idea is to replace the original IntP with an alternative scheme, while the setup for the  second stage does not change. Precisely, the two spatially separated input pulses in Fig.~1, generated by two independent lasers, are replaced with temporally separated pulses generated by a single laser, and the RPM measures their relative phase. Alice then applies a feed-forward phase shift based on the RPM measurement outcome, together with an additional phase shift $s_I$, which she chooses according to the intended output intensity. Finally, the two temporally separated pulses interfere in a one-bit-delay Mach–Zehnder interferometer (see Fig.~3 for details). 

Now, the output consists of three pulses: the middle one is the intended signal and serves as input to the second stage in Fig. 2, while the two neighboring pulses are redundant. Note that, in this alternative scheme, the delay line shown in Fig. 2 must be doubled to measure the relative phase between consecutive middle pulses. Importantly, the redundant pulses are not innocuous: they leak into the channel information about the global random phases of the gain-switched pulses. Therefore, they constitute a side channel that must be removed. A natural countermeasure is to suppress them with an intensity modulator or an optical switch, but the extinction ratio of any real intensity modulator is finite, so a residual side channel survives and must be accounted for in the security proof.

\bibliographystyle{unsrt}
\bibliography{ref}
\end{document}